\documentclass[prl,cha,onecolumn,superscriptaddress,preprintnumbers,showpacs, amsmath,amssymb]{revtex4-1}
\usepackage{bm}
\usepackage[colorlinks=true,linkcolor=blue,citecolor=blue]{hyperref}
\usepackage{amsmath}
\usepackage{amssymb}
\pdfoutput=1
\usepackage{amsthm}
\usepackage{amsfonts}
\usepackage{enumerate}
\usepackage{latexsym}
\usepackage{ifpdf}

\usepackage{graphicx}
\usepackage{makeidx}
\expandafter\ifx\csname package@font\endcsname\relax\else
 \expandafter\expandafter
 \expandafter\usepackage
 \expandafter\expandafter
 \expandafter{\csname package@font\endcsname}
 \fi
\usepackage{color}
\usepackage{times}

\begin{document}

\title{Coherent epitaxy of trilayer nickelate (Nd$ _{0.8} $Sr$ _{0.2} $)$ _{4} $Ni$  _{3}$O$ _{10}$ films by high-pressure magnetron sputtering}

\author{Jiachang Bi}
\thanks{These authors contributed equally to this work}
\affiliation{Ningbo Institute of Materials Technology and Engineering, Chinese Academy of Sciences, Ningbo, China}
\affiliation{Center of Materials Science and Optoelectronics Engineering, University of Chinese Academy of Sciences, Beijing, China}
\affiliation{School of Future Technology, University of Chinese Academy of Sciences, Beijing, China}
\author{Yujuan Pei}
\thanks{These authors contributed equally to this work}
\affiliation{School of Physics, Harbin Institute of Technology, Harbin, China}
\author{Ruyi Zhang}
\affiliation{Ningbo Institute of Materials Technology and Engineering, Chinese Academy of Sciences, Ningbo, China}
\affiliation{Center of Materials Science and Optoelectronics Engineering, University of Chinese Academy of Sciences, Beijing, China}
\author{Shaoqin Peng}
\affiliation{Ningbo Institute of Materials Technology and Engineering, Chinese Academy of Sciences, Ningbo, China}
\author{Xinming Wang}
\affiliation{Ningbo Institute of Materials Technology and Engineering, Chinese Academy of Sciences, Ningbo, China}
\author{Jie Sun}
\affiliation{Ningbo Institute of Materials Technology and Engineering, Chinese Academy of Sciences, Ningbo, China}
\author{Jiagui Feng}
\affiliation{Suzhou Institute of Nano-Tech and Nano-Bionics, Chinese Academy of Sciences, Suzhou, China}
\author{Jingkai Yang}
\email{yangjk@ysu.edu.cn}
\affiliation{Key Laboratory of Applied Chemistry, Hebei Key Laboratory of heavy metal deep-remediation in water and resource reuse, College of Environmental and Chemical Engineering, Yanshan University, Qinhuangdao, China}
\author{Yanwei Cao}
\email{ywcao@nimte.ac.cn}
\affiliation{Ningbo Institute of Materials Technology and Engineering, Chinese Academy of Sciences, Ningbo, China}
\affiliation{Center of Materials Science and Optoelectronics Engineering, University of Chinese Academy of Sciences, Beijing, China}

\date{\today}

\begin{abstract}

Rare-earth (R) nickelates (such as perovskite RNiO$_3$, trilayer R$_4$Ni$_3$O$_{10}$, and infinite layer RNiO$_2$) have attracted tremendous interest very recently. However, unlike widely studied RNiO$_3$ and RNiO$_2$ films, the synthesis of trilayer nickelate R$_4$Ni$_3$O$_{10}$ films is rarely reported. Here, single-crystalline (Nd$ _{0.8} $Sr$ _{0.2} $)$ _{4} $Ni$  _{3}$O$ _{10}$ epitaxial films were coherently grown on SrTiO$ _{3} $ substrates by high-pressure magnetron sputtering. The crystal and electronic structures of (Nd$ _{0.8} $Sr$ _{0.2} $)$ _{4} $Ni$  _{3}$O$ _{10}$ films were characterized by high-resolution X-ray diffraction and X-ray photoemission spectroscopy, respectively. The electrical transport measurements reveal a metal-insulator transition near 82 K and negative magnetoresistance in (Nd$ _{0.8} $Sr$ _{0.2} $)$ _{4} $Ni$  _{3}$O$ _{10}$ films. Our work provides a novel route to synthesize high-quality trilayer nickelate R$_4$Ni$_3$O$_{10}$ films.

\end{abstract}

\maketitle

\newpage

\section{1. Introduction}

Transition metal oxides with strong entanglements among spin, charge, orbital, and lattice degrees of freedom conceive a large number of emergent phenomena such as high-temperature superconductivity, colossal magnetoresistance, room-temperature multiferrocity, and metal-insulator transition\cite{RMP-2003-Damascelli,RMP-2001-Salamon,JPCM-1997-Ramirez,RPM-2005-Dawber,NRM-2017-Martin,RPM-1998-Imada}. One remarkable example of these transition metal oxides is the Rare-earth (R) nickelates in which metal-insulator transition, magnetic transition, and crystal structural transition were observed in perovskite nickelate RNiO$_3$\cite{PRL-2012-Chakhalian,PRL-2011-Chakhalian,PRL-2018-Chakhalian, NC-2013-Chakhalian, NC-2019-Li, PNAS-2019-Georgescu}, whereas the superconductivity ($\sim$ 15 K) presents in the infinite-layer nickelate RNiO$_2$ with doping \cite{Nature-2019-Li,APLMater-2020-Lee,NM-2020-Hepting,NanoLett-2020-Osada, NSR-2020-Gao, PRX-2021-Been,  ScientiaSinia-2021-Li,JETP-2021-Botana}. Unlike comprehensively studied RNiO$_3$ and RNiO$_2$ films\cite{PRL-2021-Li, PRB-2021-Wan, ScienceAdvances-2021-Wang, AM-2020-Wang, ChinesePhysicsB-2020-Wang, NC-2019-Liao, NC-2016-Cao}, the exploring of trilayer nickelate R$_4$Ni$_3$O$_{10}$ films is very rare\cite{APLMater-2020-Lee,CryComm-2021-Zhang}.

\begin{figure*}[]
       \includegraphics[width=0.8\textwidth]{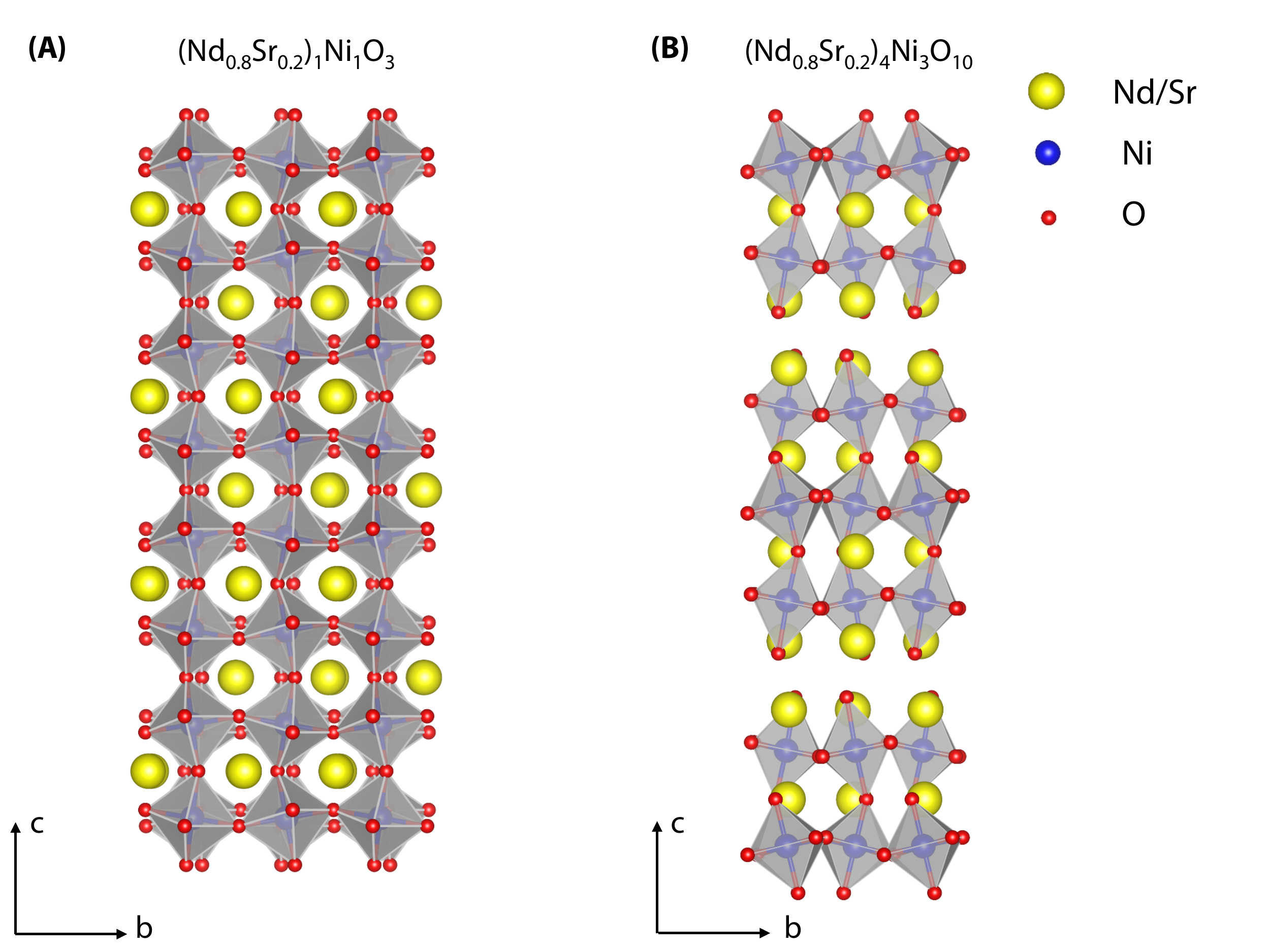}
        \caption{\label{} Schematic crystal structures of  (A) NSNO$ _{113}  $  and (B) NSNO$ _{4310}$.}
\end{figure*}

To address the above concern, we take (Nd$ _{0.8} $Sr$ _{0.2} $)$ _{4} $Ni$  _{3}$O$ _{10}$ (NSNO$ _{4310} $) films as a representative material to investigate the synthesis and the properties of R$_4$Ni$_3$O$_{10}$. The trilayer nickelate R$_4$Ni$_3$O$_{10}$ belongs to the n = 3 memeber of the Ruddlesden-Popper (RP) series of R$_{n+1}$Ni$ _{n} $O$ _{3n+1}$, which shows a rich phase diagram of novel quantum states \cite{CryComm-2021-Zhang}. For example, large hole Fermi surface can be observed in La$ _{4} $Ni$ _{3} $O$ _{10}$, which is analogous to the Fermi surface of optimally hole-doped cuprates\cite{NC-2017-Li}. Besides this property, metal to metal transition, charge order, intertwined charge, and spin density wave were also shown in R$_{4} $Ni$_{3} $O$ _{10}$ \cite{JSolidState-1995-Zhang,JAP-2000-Carvalho,NC-2020-Zhang,PRB-2020-Huangfu,PRB-2020-Li,PRB-2020-Rout}. Additionally, in the reduced phase of R$ _{4} $Ni$ _{3} $O$ _{10}$ (such as R$_{4} $Ni$_{3} $O$_{8}$, R = La, Nd, Pr), charge/spin stripes, charge-stripe fluctuations, and orbital polarization were seen \cite{PRB-2021-Hao,PNAS-2016-Zhang,PRL-2019-zhang,NP-2017-Zhang}. It is noted that the synthesis of high-quality bulk single crystals has been a big challenge\cite{JAP-2000-Das,JAC-2004-Zinkevich} and only successful in two groups \cite{PNAS-2016-Zhang,NP-2017-Zhang,PRL-2019-zhang,PRM-2020-Zhang,PRB-2020-Huangfu}.

\begin{figure*}[]
        \includegraphics[width=0.9\textwidth]{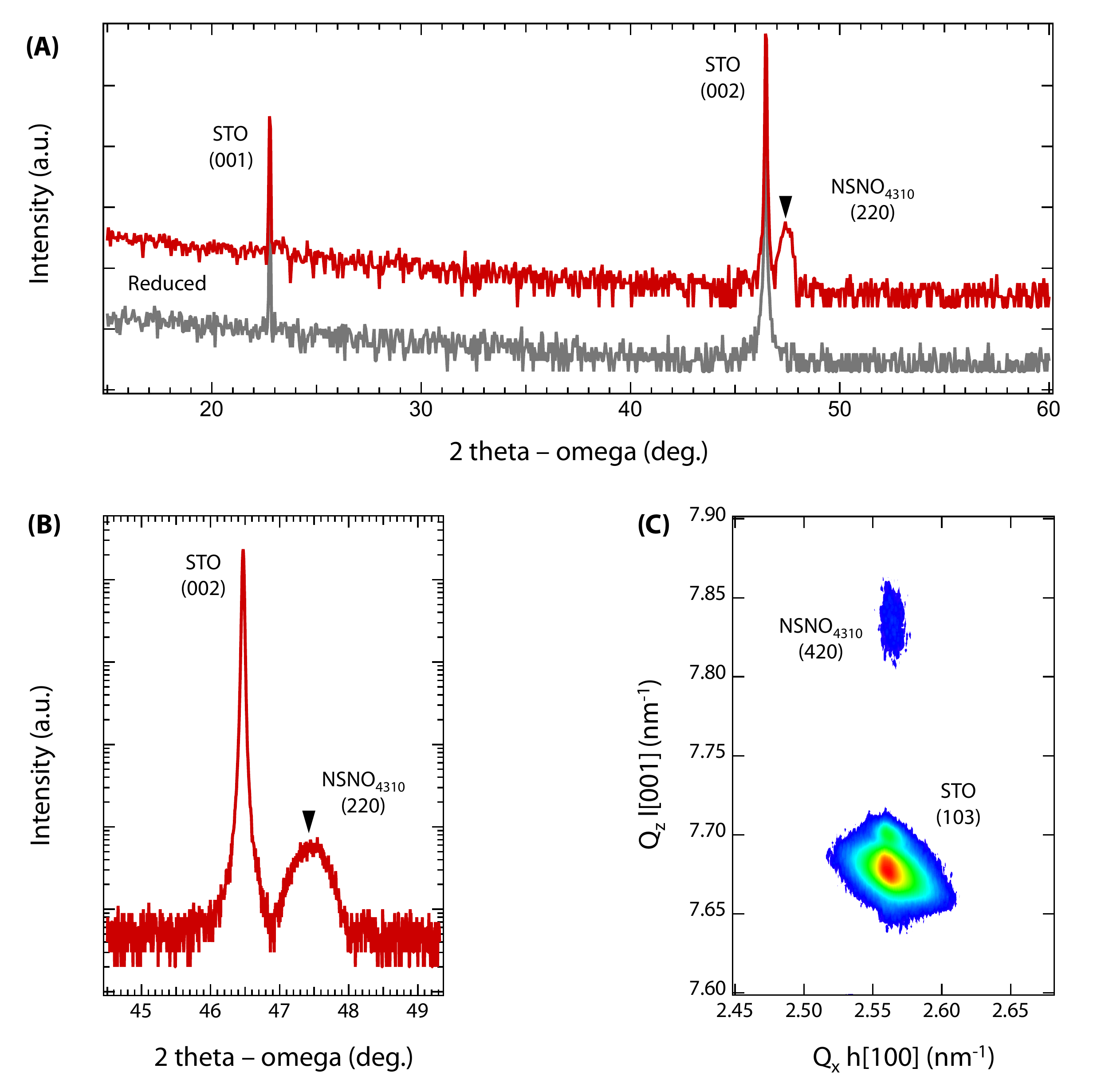}
        \caption{\label{} (A) Wide-range 2$\theta$-$\omega$ scan of NSNO$ _{4310} $ (red) and NSNO$ _{438} $ (grey) films along the STO (00L) diffraction. The black triangle indicates the (220) peak of  NSNO$ _{4310} $ films. (B) Expanded view of Figure (A) near the STO (002) peak. (C) RSM pattern of NSNO$ _{4310} $ films (top) around asymmetric STO (103) diffraction (bottom).}
\end{figure*}

Here, high-quality NSNO$ _{4310} $ epitaxial films were synthesized on SrTiO$ _{3} $ (STO) substrates by high-pressure magnetron sputtering. The crystal and electronic structures of NSNO$ _{4310} $ films were characterized by high-resolution X-ray diffraction (HRXRD) and X-ray photoemission spectroscopy (XPS). Electrical transport measurements reveal a metal-insulator transition (MIT) near 82 K and negative magnetoresistance. Our work provides a novel route to synthesize high-quality trilayer nickelate R$_4$Ni$_3$O$_{10}$ films.

\begin{figure*}[]
        \includegraphics[width=0.8\textwidth]{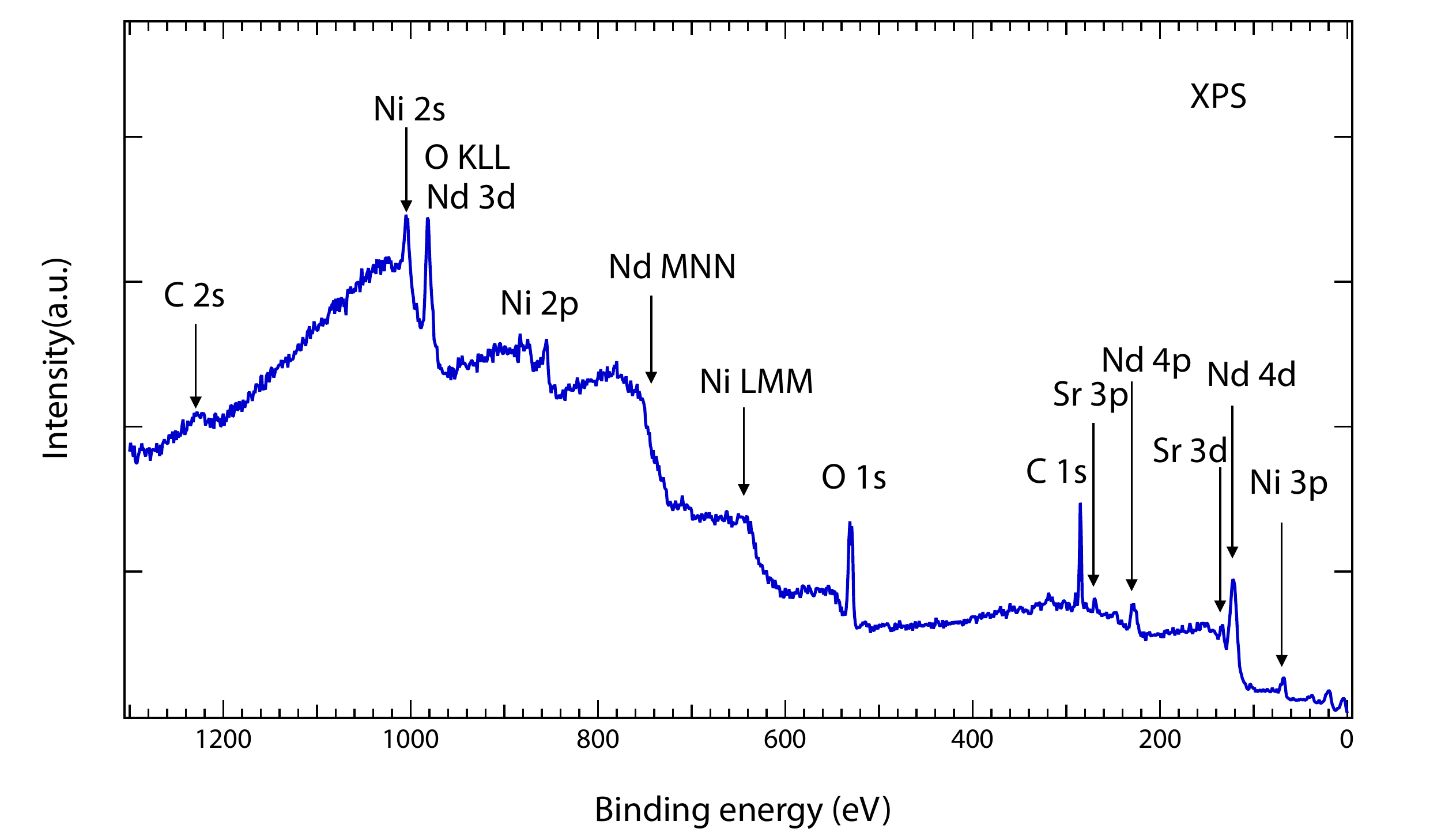}
        \caption{\label{} XPS spectra of NSNO$ _{4310} $ films on STO single-crystalline substrates at room temperature.}
\end{figure*}

\section{2. Materials and methods}

2-inch Nd$ _{0.8} $Sr$ _{0.2} $NiO$ _{3} $ polycrystalline targets were synthesized by a standard solid-state reaction method with initial reactants consist of Nd$ _{2}$O$ _{3} $(Aladdin, 99.99$\% $), SrCO$ _{3} $(Aladdin, 99.99$\% $), and NiO(Aladdin, 99.99$\% $). The mixture was calcined at 1350 $^\circ$C for 8 h and reground for 6 h. Then the powders were pressed into a 2-inch target with 30 MPa and annealed at 1300$^\circ$C for 2 h. The high-quality NSNO$ _{4310} $ thin films ($\sim$ 20 nm) were synthesized by high-pressure radio frequency (RF) magnetron sputtering (home-made) with a 2-inch target and the O$_2$ (purity of 99.999\%) reactive gas \cite{ACS-2021-Zhang, APLM-2021-Zhang}. Before growth, the base vacuum pressure was $\sim$ 3 $\times$10$^{-8}$ torr. During growth, O$_2$ pressure was kept at 0.02 Torr with a gas flow of 1.8 sccm, and the substrate temperature was held at 550 $^\circ$C. The power of the RF generator was kept at 60 W. To ensure the uniformity of the films, the heating stage was rotating at a speed of 5 rams/min during growth. After growth, the  films were cooled down to room temperature at 25$^\circ$C per minute in the 0.02 Torr O$_2$ atmosphere. A topochemical reduction process transforms NSNO$ _{4310} $ films into NSNO$ _{438} $ films. The NSNO$ _{438} $ films and reducing agent CaH$ _{2} $ were sealed in an evacuated quartz tube and heated at 280 $^\circ$C for 6 hours.

The crystal structure of NSNO$ _{4310} $ films was characterized by the High-resolution X-ray diffractometer (Bruker D8 Discovery) with the Cu K$ _{\alpha} $ source ($\lambda$ = 1.5405 \AA). The 2$\theta$-$\omega$ scans and asymmetrical reciprocal space mappings were performed to reveal the coherent growth and lattice parameters of films. The electronic structure of the films was probed by PHI 5000 Versa Probe x-ray photoelectron spectroscopy (XPS) at an acceptance angle of 45$ ^{\circ} $ for the analyzer (using monochromated Al K$\alpha$ radiation, h$\upsilon$ = 1486.6 eV). The electrical properties were measured by Physical Property Measurement System (PPMS from Quantum Design) with four-point probe measurements. 

\section{3. Results and discussions}

Figure 1 shows the crystal structures of NSNO$ _{113}$ and NSNO$ _{4310}$. NSNO$ _{113} $ has a symmetry of orthorhombic structure with the lattice parameters \textit{a} = 5.39 \AA, \textit{b} = 5.38 \AA,\textit{c} = 7.61 \AA ~(the pseudocubic lattice parameter is $\sim$ 3.81 \AA). It is noted that the replacement of Nd by Sr atoms via doping has little impact on the symmetry of crystal structure and the lattice parameters \cite{PRB-1995-Alonso}. In contrast, the NSNO$ _{4310}$ has a monoclinic symmetry (space group \textit{P2$ _{1} $/a}) with the lattice parameters \textit{a} = 5.365 \AA, \textit{b} = 5.455 \AA, \textit{c} = 27.418 \AA, and $\beta $ = $ 90.31^{\circ}$ at room temperature \cite{JSolidStateChem-2000-Olafsen,PRB-2020-Li}.

 \begin{figure*}[]
        \includegraphics[width=0.8\textwidth]{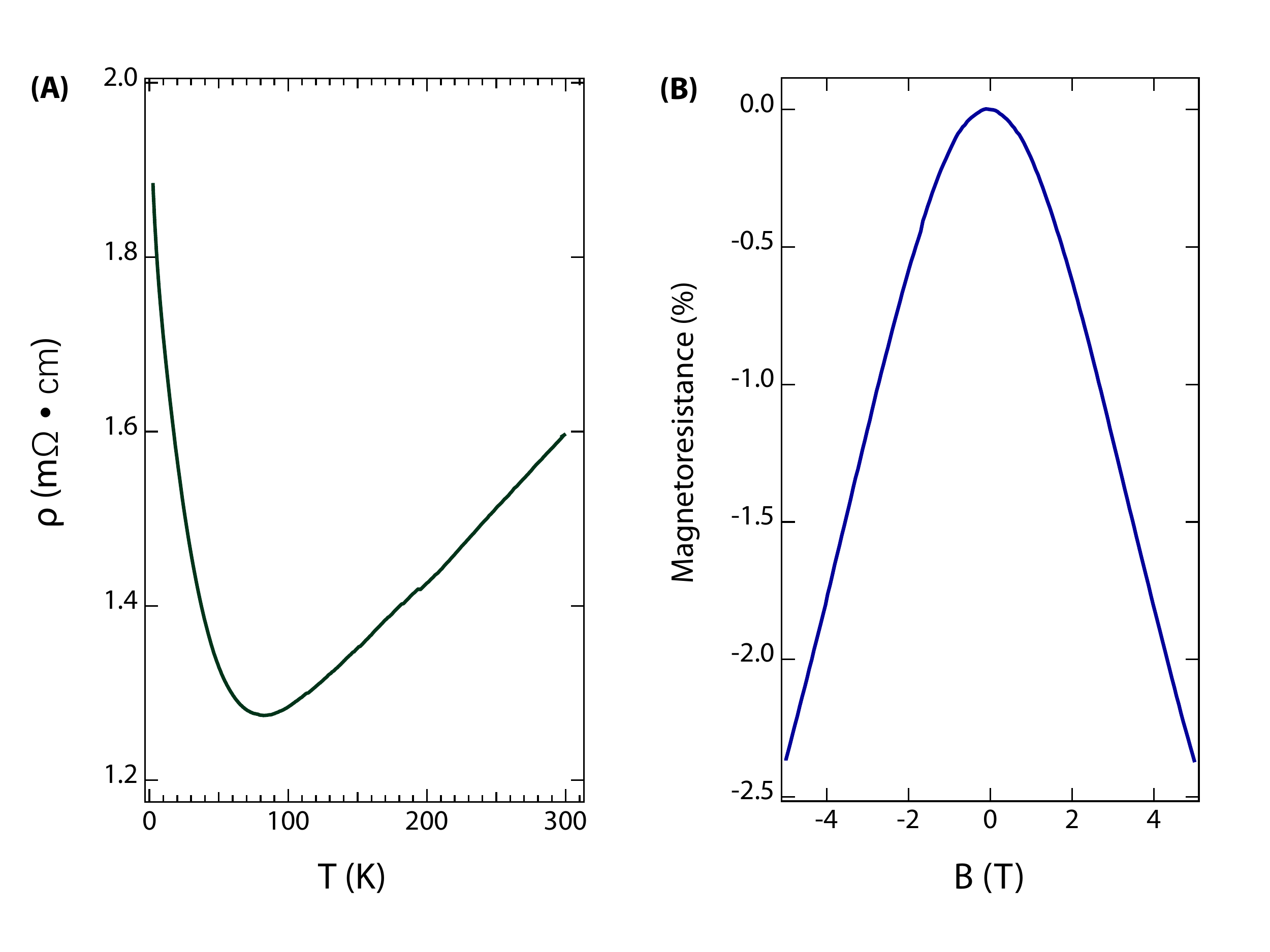}
        \caption{\label{} Transport properties NSNO$ _{4310} $ film. (a) Resistivity versus temperature $ \rho  $ (T) curves from 2.5 K to 300 K. (b) Magnetoresistance $ \Delta\rho / \rho_{0}$ at 2.5 K.}
\end{figure*}

First, we performed HRXRD to characterize the crystal structure of NSNO$ _{4310}$ films. The wide-range 2$\theta$-$\omega$ scans in Figure 2 (A) show that the (220) diffraction peak of NSNO$ _{4310}$  is along with the (00L) diffraction peaks of STO substrates without any secondary phases, whereas the (110) peak disappears due to the symmetry\cite{JSolidStateChem-2000-Olafsen}, indicating that NSNO$ _{4310}$ film is highly textured on STO substrates. The amplified view near (002) diffraction of STO substrates (see Figure 2 (B)) shows a film peak $\sim $ $ 47.46^{\circ}$, corresponding to a (110) layer spacing ($\sim$ 3.828 \AA) of  NSNO$ _{4310}$ films, which agree well with the results in the literature \cite{JSolidStateChem-2000-Olafsen,APLMater-2020-Lee}. Reducing NSNO$ _{4310}$ films by CaH$ _{2} $ can generate (Nd$ _{0.8} $Sr$ _{0.2} $)$ _{4} $Ni$  _{3}$O$ _{8} $ (NSNO$ _{438} $) with diffraction angle at $\sim $ $ 46.47^{\circ}$, which is too close to the STO (002) peak near  $ 46.47^{\circ}$ to be seen \cite{JSolidState-1992-Lacorre, APLMater-2020-Lee}. Hence, no significant feature is observed in Figure 2 (A) for NSNO$ _{438}$ films. 
 
Next, to characterize the in-plane lattice parameters of NSNO$ _{4310}$ films, we measured the reciprocal space mapping (RSM). Figure 2 (C) shows the RSM recorded around asymmetric STO (103) diffraction. The reciprocal lattice vectors Q$_{x} $ and Q$_{z}$ correspond to the in-plane (100) and out-of-plane (001) directions of STO single crystalline substrates. The lower  diffraction peak can be assigned to the STO, whereas the upper diffraction peak corresponds to the epitaxial NSNO$ _{4310}$. The epitaxial relationship can be expressed as, NSNO$_{4310}$[1-10]//STO[100] and NSNO$_{4310} $[001]//STO[010] for in-plane direction and NSNO$ _{4310} $(110)//STO(001) for out-of-plane direction. As revealed by RSM, NSNO$ _{4310} $ film shares the same  Q$ _{x}$ value with that of STO substrates, indicating a coherent epitaxial growth for NSNO$ _{4310} $ film with an in-plane (1-10) layer spacing $\sim$ 3.905 \AA. From RSM, it is estimated that the (110) layer spacing is $\sim$ 3.828 \AA, agreeing very well with the value extracted from the (220) diffraction peak of NSNO$ _{4310} $ films. Thus, the RSM data further confirms the high-quality of epitaxial NSNO$ _{4310}$ films. 

To further characterize the chemical composition of the NSNO$ _{4310} $ film, the XPS characterization was carried out. Figure 3 shows a wide-energy core-level XPS spectrum from 0 to 1300 eV. As seen, besides adsorbed carbon on the NSNO$ _{4310} $ film surface, no detectable impurity signal is observed.

Figure 4 (A) shows the temperature-dependent resistivity of NSNO$ _{4310}$ films from 2.5 K (1.88 $ m \Omega \cdot cm   $) to 300 K (1.59 $ m \Omega \cdot cm   $). There is a metal-insulator transition at $\sim $ 82 K. It is noted that the resistivity remains small at the range of 2.5 - 300 K, indicating a metallic behavior. Magnetoresistance (MR = $ \Delta \rho /\rho _{0} \times 100\%$) at 2.5 K was also measured (see Figure 4 (B)). The negative magnetoresistance of NSNO$ _{4310}$ films has also been observed in the bulk \cite{ScienceChina-2021-Li}, resulting from the weak localization at low temperature and being different from the electrical behavior of La$ _{4} $Ni$ _{3} $O$ _{10}$\cite{MMM-2020-Kumar} and Pr$ _{4} $Ni$ _{3}  $O$ _{10}$\cite{PRB-2020-Huangfu}.

\section{4. Conclusion}

In this work, high-quality (Nd$ _{0.8} $Sr$ _{0.2} $)$ _{4} $Ni$  _{3}$O$ _{10} $ films have been synthesized on SrTiO$ _{3} $ by high-pressure magnetron sputtering. The crystal and electronic structures of (Nd$ _{0.8} $Sr$ _{0.2} $)$ _{4} $Ni$  _{3}$O$ _{10} $ films are characterized by XRD and XPS. The electrical transport measurements reveal a metal-insulator transition and negative magnetoresistance. Our work provides a novel route to synthesize high-quality trilayer nickelate R$_4$Ni$_3$O$_{10}$ films.

\section{Acknowledgments}

The authors deeply acknowledge the insightful discussions
with Liang Wu. This work is supported by the National Natural Science Foundation of China (Grant Nos. 11874058 and U2032126), the Pioneer Hundred Talents Program of the Chinese Academy of Sciences, the Natural Science Foundation of Zhejiang Province, the Beijing National Laboratory for Condensed Matter Physics, and the Ningbo Science and Technology Bureau (Grant No. 2018B10060). This work is partially supported by the Youth Program of the National Natural Science Foundation of China (Grant No. 12004399), China Postdoctoral Science Foundation (Grant No. 2018M642500), and Postdoctoral Science Foundation of Zhejiang Province (Grant No. zj20180048).

\newpage

\end{document}